\documentclass[%
reprint,
showpacs,preprintnumbers,
nofootinbib,
 amsmath,amssymb,
aps,
prb,
floatfix,
]{revtex4-1}

\usepackage{graphicx}
\usepackage{dcolumn}
\usepackage{bm}
\usepackage{hyperref}
\usepackage{color}

\begin{document}


\title{Strain distribution in polycrystals: Theory and Application for Diffraction Experiments}

\author{\'Ad\'am Tak\'acs}
\email{takacs.adam@wigner.mta.hu}
\affiliation{Department of Materials Physics, E\"otv\"os University, P\'azm\'any P\'eter s\'et\'any 1/A, H-1117 Budapest, Hungary;
}%
\author{G\'eza Tichy}%
 \email{tichy@caesar.elte.hu}
\affiliation{Department of Materials Physics, E\"otv\"os University, P\'azm\'any P\'eter s\'et\'any 1/A, H-1117 Budapest, Hungary;
}%
\author{P\'eter Dus\'an Isp\'anovity}
\email{ispanovity@metal.elte.hu}
\affiliation{Department of Materials Physics, E\"otv\"os University, P\'azm\'any P\'eter s\'et\'any 1/A, H-1117 Budapest, Hungary;
}%

\date{\today}

\begin{abstract}
Randomly textured polycrystalline materials of constituents with highly anisotropic nature of grains can be considered globally isotropic. In order to determine the isotropic properties, like elasticity or conductivity, we propose a theory for averaging the coefficients of the corresponding tensors unifying Voigt's, Reuss' or other self-consistent homogenization theories. We apply the method to determine elastic moduli of untextured polycrystals with arbitrary crystal structures, recovering experimental data with high precision for cubic materials. We show that the average moduli can be used to predict analytically stress and strain states inside individual grains as proven by the comparison with neutron diffraction measurements. Finally, we discuss a few possible generalizations for textured materials for further applications.
\end{abstract}

\pacs{46.25.-y, 61.05.C-, 61.05.F-, 61.14.Dc, 61.50.Ah, 62.20.Dc, 68.55.Jk}


\maketitle

\section{Introduction}

It is a frequently encountered phenomenon in nature that macroscopic properties of a system are isotropic, although its microscopic constituents have anisotropic character. Common examples for such behavior are polycrystals with anisotropic grains or molecular gases. To predict macroscopic properties of the whole system it is not enough to have knowledge on the anisotropic subparts, but some information is needed about their disorder. A common idea to perform the averaging procedure is \textit{self-consistency} when one assumes that the overall response is identical to the average response of the constituents. This technique is widely used in condensed matter physics \citep{dewit2008elastic,Kanouté2009,Kroner1958,CLAUSEN19983087,2015GeoJI.203..334G,Wu2016,MAKARIAN201617659,MILED201696,2016CMAME.310..749V,CAZACU2017,2017arXiv170510018C,BUI2017} and also in molecular dynamics \citep{salam2010molecular}.

This paper is concerned with the elastic properties of polycrystalline materials. Nowadays, the most widespread method for such problems is the crystal plasticity finite-element (CPFE) method~\citep{Segurado20133, Engler20052241, Raabe20024379, VanHoutte2002359, xue2000modeling, BECKER19952701, SPAHN2014871, ROTERS20101152, ma10101172}. It provides a full-field solution for the elasto-plastic problem given that all the texture information is available~\citep{0965-0393-11-3-201,ARSENLIS20041213,VITEK200431}. The variational formulation of this method guarantees the equilibrium of the forces and the compatibility of the displacements during the whole course of deformation. CPFE simulations have been applied to model texture evolution~\citep{LI2005207,VANHOUTTE2006634,LEE2007263}, grain boundary migration~\citep{WEI20042587,BATE2005199}, micron-scale plastic deformation~\citep{CHEONG20071757,LIU2007130}, high temperature deformation phenomena and recrystallization~\citep{BOWER20041289,doi:10.1080/14786430802502575}; just to mention a few examples. Despite their extreme versatility, the usability of CPFE simulations is often limited due to their very large number of degrees of freedom that provides strict upper limit to the specimen volume that can be modeled. In addition, it requires the full knowledge of the grain structure, which is not always available.

In this paper we aim at providing a theoretical explanation for results of neutron diffraction measurements performed on an elastically loaded untextured material. Diffraction measurements usually do not know about the exact grain structure, they can only measure some statistical properties (like average or scatter) of the grain sizes. We, therefore, focus on developing a self-consistent method to predict the internal strain inside grains with different orientations in an untextured sample. To this end, we first review a general method to perform angular averaging of tensors like elastic constants of an untextured polycrystal. Then, we provide a new homogenization method that is similar to Voigt's~\citep{voigt1887}, Reuss'~\citep{reuss1929} or Hill's~\citep{hill2002} model, but is more general. This is followed by an analytical mean-field calculation for a spherical grain embedded in an elastically deformed isotropic virtual medium representing the untextured surrounding polycrystal. Finally, the predictions of this model are compared to the results of the diffraction measurements where the orientation dependent diffraction patterns were measured during mechanical deformation. The paper concludes with an outlook on possible generalizations and applications.

\section{Angular averaging of tensors}\label{sec:tensor}

Tensors are used to represent anisotropic but linear connection between physical quantities. For example the anisotropic electrical conductivity is represented by a $3\!\times \!3$ matrix or the elastic stiffness tensor by a $3\!\times \!3\!\times \!3\!\times \!3$ matrix. In the case when isotropy is caused by randomly oriented subparts, the isotropic connection between physical quantities can be predicted by averaging the tensor with respect to the direction of a corresponding base. Equivalently, one can predict the isotropic physical quantities by rotating randomly the tensor of the anisotropic connection.

The general linear connection $\bm{\mathcal{C}}$ in 3 dimension between $\bm{A}$ and $\bm{B}$ matrices with $N$ indices reads as
\begin{equation}
A_{i_1 \dots i_N} = \mathcal{C}_{i_1 \dots i_N;j_1 \dots j_N} B_{j_1 \dots j_N},
\end{equation}
where $i,j=1,2,3$ and we use Einstein's notation for summation. To average the tensor connection $\bm{\mathcal{C}}$, one rotates it and performs the averaging over the coefficients with respect to the angles of rotation. The rotation can be formulated using the rotation matrix $R_{ij}=f^{(j)}_i$, where $\bm f^{(j)}$ is the $j^\text{th}$ base vector of the rotated frame, and $i,j=1,2,3$, thus
\begin{equation}
\begin{split}
\mathcal{C}'_{i_1 \dots i_N;j_1 \dots j_N}=&R_{i_1k_1}\dots R_{i_Nk_N}\mathcal{C}_{k_1 \dots k_N;l_1 \dots l_N}\cdot \\ 
&R^\text{T}_{l_1j_1} \dots R^\text{T}_{l_Nj_N}.
\end{split}
\end{equation}
The directional average [denoted by $\overline{(\dots)}$] is performed over the angles characterizing the frame $\bm f^{(i)}$. Since $\bm{\mathcal{C}}$ is independent from these angles, the result can be expressed by the direction averaged products of the bases
\begin{equation}
\overline{\mathcal{C}}_{i_1 \dots i_N;j_1 \dots j_N} = \overline{f^{(k_1)}_{i_1}...\,f^{(k_N)}_{i_N}f^{(j_{1})}_{l_{1}}...\,f^{(j_{N})}_{l_{N}}}\, \mathcal{C}_{k_1 \dots k_N;l_1 \dots l_N},
\end{equation}
where the direction average is defined as
\begin{equation}\label{eq:rotational_average}
\overline{f^{(i_1)}_{j_1}...\,f^{(i_{2N})}_{j_{2N}}}=\int d\zeta_1...\,d\zeta_{2N} f^{(i_1)}_{j_1}...\,f^{(i_{2N})}_{j_{2N}}.
\end{equation}
Here, $\zeta_i$ denotes the integration variables of the rotation, and includes the Jacobian determinant divided by the total measure of the angles for normalizing. If one chooses the Euler angles (denoted by $\alpha$, $\beta$, and $\gamma$) to characterize the basis orientation then the normalized integration variables read as $d\zeta_1d\zeta_2d\zeta_3 =d\alpha d\beta d\gamma\sin\beta/8\pi^2$. By definition $\overline{f^{(i)}_j}=0$ and similarly vanishes for odd number of factors. For the product of two base vectors, the result is $\overline{f^{(i)}_jf^{(k)}_l}=1/3 \, \delta_{ij} \delta_{kl}$, which leads to a simple rule for rank-2 matrices:
\begin{equation}\label{eq:3x3_averging}
\overline{\bm{\mathcal{C}}}=\frac{1}{3} \bm{I}\,\text{Tr}\, \bm{\mathcal{C}},
\end{equation}
with $\bm I$ being the identity matrix. For the average of products of four basis vectors one obtains
\begin{equation}
\label{eq:fourmean}
\begin{split}
&\overline{f^{(i)}_pf^{(j)}_qf^{(k)}_rf^{(l)}_s}= \\
&\quad
\begin{cases}
\frac{1}{5} , & \text{if } \delta_{ijkl},\text{ } \delta_{pqrs}\\
\frac{1}{15} , & \begin{split}&\text{if } \delta_{ijkl},\text{ } \delta_{pq},\text{ }  \delta_{rs},\text{ }p\neq r,\\& \text{or } \delta_{ij}, \text{ } \delta_{kl},\text{ } i\neq k, \text{ } \delta_{pqrs}\end{split}\\
\frac{2}{15} , & \text{if } \delta_{ij}, \text{ } \delta_{kl},\text{ } i\neq k, \text{ } \delta_{pq}, \text{ } \delta_{rs},\text{ } p \neq r\\
-\frac{1}{30} , & \begin{split}&\text{if } \delta_{ij}, \text{ } \delta_{kl},\text{ } i\neq k, \text{ } \delta_{pr}, \text{ } \delta_{qs},\text{ } p \neq q ,\\&
\text{or } \delta_{ik}, \text{ } \delta_{jl},\text{ }i\neq j, \text{ } \delta_{pq}, \text{ } \delta_{rs},\text{ } p \neq r \end{split} \\
0 , & \text{otherwise,}\\
\end{cases}
\end{split}
\end{equation}
where $\delta_{ijkl}:=1$, if $i=j=k=l$ or 0 otherwise. For the averaged factors of the higher ranked tensors see Appendix B in Ref.~\citep{salam2010molecular}.\footnote{Calculations in Appendix B of Ref.~\citep{salam2010molecular}, contrary to our case, yield non-vanishing results for the product of odd number of base vectors, due to the assumption of parity transformation, which does not hold in the present case.}

This averaging procedure can be generalized for the case of textured microstructure. In this case, the rotational average has to be performed with using specific weight-factors in Eq.~\eqref{eq:rotational_average}, see, e.g., Ref.~\citep{Kube2016}.

\section{Elastic Coefficients of Virtual Medium}\label{sec:virtual_medium}
Untextured polycrystals are homogeneous and isotropic, so, one can apply the above mentioned averaging procedure to predict the isotropic properties of the specimen. For example, anisotropic relation inside the grains between the current density $\bm J$ and the electric field $\bm E$ is $J_i=\Sigma_{ij}E_j$, where $\bm{\Sigma}$ is the electrical conductivity. From Eq.~\eqref{eq:3x3_averging} the direction averaged connection for polycrystals is approximately
\begin{equation}
\bm{\bar{\Sigma}}=\frac{1}{3}\,\bm{I}\, \text{Tr}\,\bm{\Sigma},
\end{equation}
therefore the conduction can be expressed with simple scalar equation $J_i=\bar{\Sigma}E_i$. This result is used, for instance, to predict permeability in monodisperse materials or magneticity in alloys \cite{2016JAG...127...82Y,doi:10.1177/1081286512458109}. 

The elastic coefficients are described by the elastic stiffness tensor $\bm C$ with four indices, which describes linear relationship between strain and stress
\begin{equation}
\sigma_{ij}=C_{ijkl}\varepsilon_{kl}.
\end{equation}
The inverse of the stiffness tensor is called compliance tensor $\bm S$, which fulfills
\begin{equation}
\varepsilon_{kl}=S_{klij} \sigma_{ij}.
\end{equation}

The direction average of a rank-4 matrix can be done by using the average of the base vector products from Eq.~\eqref{eq:fourmean}. The non-zero coefficients of the averaged stiffness tensor in sextic notation for a material with cubic symmetry are
\begin{equation}\label{eq:C_sextic_rotated}
\begin{split}
\overline{C}_{11} = C_{11}-\frac{2}{5}(C_{11}-C_{12}-2C_{44}),\\
\overline{C}_{12} = C_{12}+\frac{1}{5}(C_{11}-C_{12}-2C_{44}),\\
\overline{C}_{44} = C_{44}+\frac{1}{5}(C_{11}-C_{12}-2C_{44}).
\end{split}
\end{equation}
The averaged coefficients indeed satisfy the isotropic condition, that is, the degree of isotropy $a$, defined as
\begin{equation}
a=\frac{2C_{44}}{C_{11}-C_{12}},
\end{equation}
equals to 1. It is important to point out that the similarly performed averaging operation for the compliance tensor leads to a different result in the sense $\bm{\overline{C}}\neq(\bm{\overline{S}})^{-1}$.

\begin{figure}[h]
\centering
    \includegraphics[width=0.5\textwidth]{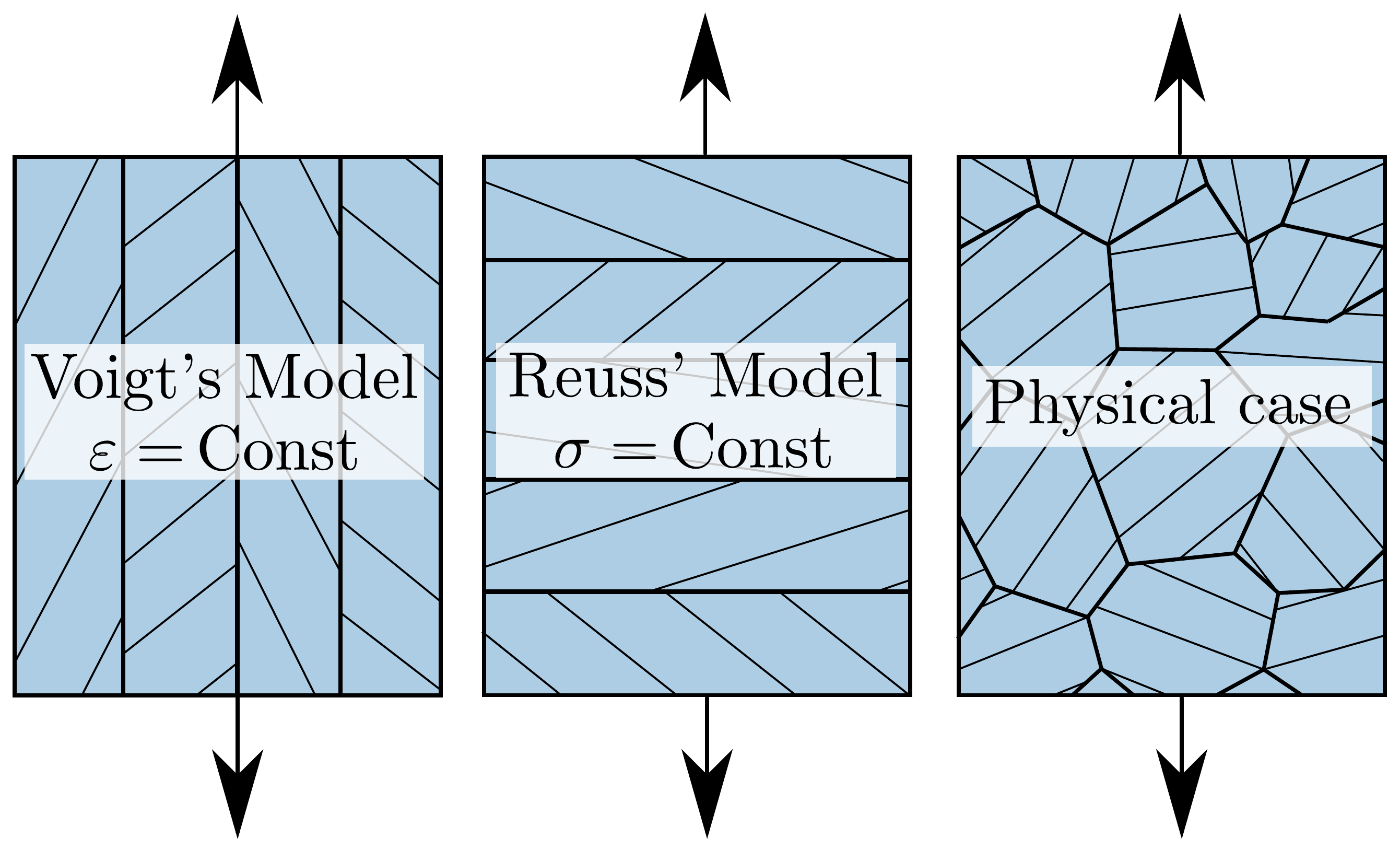}
    \caption{\label{fig:Voigt_Reuss_physical} The first two panels show the assumed arrangement of the grains and the corresponding Voigt's and Reuss' averaging schemes. The third panel represents the real untextured polycrystal systems which one has to combine both model. }
\end{figure}
The introduced direction averaging process assumes randomly oriented grains inside polycrystals, yet, as said above, it is not irrelevant, whether one averages $\bm{C}$ or $\bm{S}$. To tackle this issue it is important to study the physical arrangement of the grains. The first two panels of Fig.~\ref{fig:Voigt_Reuss_physical} illustrate the two most common assumptions, the Voigt and the Reuss set-up. In Voigt's model~\cite{voigt1887} the strain is homogeneous, therefore the average process would be executed over the stiffness tensor (briefly $\overline{\bm{C}}$). This assures compatibility, but not necessary equilibrium in a polycrystal. However, in Reuss' model~\cite{reuss1929} the stress is homogeneous and the average would run over the compliance (briefly $\overline{\bm{S}}$). This assumes equilibrium, but not necessary compatibility. In a physical situation, like the one sketched in the right panel of Fig.~\ref{fig:Voigt_Reuss_physical}, the elastic coefficients of the perfectly untextured polycrystalline material must lie between these two limits. Our aim is now to introduce a suitable model for such case, but stay as general as possible.

To interpret models of Voigt and Reuss we first introduce the generalized $p$-mean of quantity $X$ as
\begin{equation}
\langle X \rangle_p=\left( \frac{1}{N}\sum^{N}_{i=1}X^p_i \right)^{1/p}.
\end{equation}
With this notation Voigt's model corresponds to $\overline{\bm{C}} = \langle \bm C \rangle_1$ (arithmetic mean) and Reuss' to $\overline{\bm{S}} = \langle \bm C \rangle_{-1}$ (harmonic mean). Since generalized mean is a monotonic function of the exponent $p$, here we choose $p\rightarrow0$ for calculating the isotropic elastic constants (geometric mean) which is between the previous limits. This choice is also motivated by Hill~\citep{hill2002}, who showed that Voigt's and Reuss' assumptions lead to the least upper bound and the greatest lower bound for the elastic coefficients. Hence, in our calculations the total averaged stiffness and compliance tensors ($\overline{\bm{C}}_\text{av}$ and $\overline{\bm{S}}_\text{av}$, respectively) are formally expressed by the geometric mean of the Voigt's and Reuss' model
\begin{eqnarray}\label{eq:geomean}
\overline{\bm{C}}_\text{av} = \sqrt{\overline{\bm{C}}\,\overline{\bm{S}}^{-1}},\\
\overline{\bm{S}}_\text{av} = \sqrt{\overline{\bm{C}}^{-1}\,\overline{\bm{S}}} = \overline{\bm{C}_\text{av}}^{-1}. \label{eq:geomean2}
\end{eqnarray}
In the rest of this paper the term \textit{virtual medium} will refer to an isotropic material with the obtained averaged elastic properties $\overline{\bm{C}}_\text{av}$. It is noted that there are several other models to predict the average elastic properties of polycrystals (see, e.g., the reviews in Refs.~\citep{dewit2008elastic,Kanouté2009,Man2011}). The advantage of our approach is its generality and the fact that it is irrelevant whether the averaging is performed on $\bm C$ or $\bm S$ [see Eq.~\eqref{eq:geomean2}].
 
To verify the correctness of the averaging procedure, it was applied to different cubic crystals (although it can be performed for any crystal structure). Columns 2, 3 and 4 in Table~\ref{tab:coeff1} show the experimentally measured elastic coefficients of the single crystalline form of these materials. Column 5 contains the degree of anisotropy $a$, defined above, value of 1 corresponding to isotropy. Columns 6, 7 and 8 give the averaged elastic coefficients from Eq.~\eqref{eq:geomean}. It is evident from the table that, averaging indeed changed the coefficients significantly which now satisfy the isotropic condition. Columns 7 and 8 correspond to the Lam\'{e} coefficients of the isotropic virtual medium (denoted by $\bar{\lambda}$ and $\bar{\mu}$), whereas the last two columns contain Lam\'{e} coefficients measured on real untextured polycrystals (denoted by $\lambda$ and $\mu$). As it is seen the deviation between measured and predicted values is only about a few percent, justifying the chosen averaging method.
\begin{table*}
\begin{ruledtabular}
\begin{tabular}{c|cccc|ccc|cc}
\multicolumn{1}{c}{Materials} & \multicolumn{4}{c}{Measured Single Crystal} & \multicolumn{3}{c}{Calculated Virtual Medium} & \multicolumn{2}{c}{Measured Polycrystal}\\
 &   $C_{11}$ &  $C_{12}$ & $C_{44}$ & $a$ & $\bar{C}_{\text{av},11}$ & $\bar{C}_{\text{av},12}\equiv \bar{\lambda}$ & $\bar{C}_{\text{av},44}\equiv \bar{\mu}$ &  $\lambda$ & $\mu$ \\ \colrule
 Ag &          12.4 &  9.30  & 4.60 & 2.90  & 14.27  &  8.37 & 2.95 & 8.130 & 2.96\\
 Al &          10.8  & 6.10  & 2.90 & 1.20   & 11.23 &  5.89 & 2.67 &  5.910 & 2.61 \\
 Au &          18.6  & 15.7  & 4.20 & 2.90   & 20.34 &  14.9 & 2.72 &  15.26 & 2.75  \\
 Na &          0.73  & 0.63  & 0.42 & 8.30   & 0.890 &  0.55 & 0.17 &  0.370 & 0.20 \\
 Cu &          16.8  & 12.1  & 7.50 & 3.20   & 19.92 &  10.6 & 4.66 &  10.24 & 4.74 \\
 $\alpha$-Fe & 23.7  & 14.1  & 11.6 & 2.40   & 28.12 &  11.9 & 8.11 &  11.08 & 8.19 \\
 Mo &          46.0  & 17.6  & 11.0 & 0.78 & 43.30 &  18.9 & 12.2 &  17.77 & 12.7  \\
 Ni &          24.7 & 14.7  & 12.5 & 2.50  & 29.54 &  12.3 & 8.62 &  12.48 & 8.23  \\
 Pb &          5.00  & 4.20  & 1.50 & 3.70   & 5.630 &  3.89 & 0.87 &  3.670 & 0.85 \\
 W &           50.1  & 19.8  & 15.1 & 1.00  & 50.10 &  19.8 & 15.1 &  20.58 & 16.2 \\ 
 Diamond &    108.0  & 12.5  & 57.6 & 1.20   & 115.2 &  8.64 & 53.3 &  8.660 & 53.3 \\
 Si &          16.6  & 6.40  & 8.10 & 1.60   & 18.71 &  5.35 & 6.68 &  5.360 & 6.66 \\
 MgO &        28.92 & 8.80 & 15.46 & 1.54  & 32.851 & 6.829 & 13.01 & 7.384 & 12.9 \\
\end{tabular}
\end{ruledtabular}
\caption{\label{tab:coeff1} Columns 2-4 and 5 show the measured elastic coefficients of some cubic materials in $10^{10}$ Pa and their degree of anisotropy, respectively Ref.~\citep{courtney2005mechanical,vahldiek1968anisotropy}. Columns 6-8 show the calculated isotropic elastic coefficients of virtual medium from Eq.~\eqref{eq:geomean}. The last two columns show experimentally measured Lam\'{e} coefficients of polycrystals in $10^{10}$ Pa from Ref.~\citep{courtney2005mechanical,vahldiek1968anisotropy}. Comparison of columns 7-9 and 8-10 shows that our model is able to predict the elastic coefficients with only a few percent of deviation.}
\end{table*}

In case of textured materials, as it was mentioned at the end of Sec.~\ref{sec:tensor}, the average could be calculated using weight factors in Eq.~\eqref{eq:C_sextic_rotated} representing the anisotropy of the grain structure (that is, texture).

\section{Stress Inside Anisotropic Grains}\label{sec:InsideStress}

\begin{figure}[h]
\centering
  \includegraphics[width=0.9\linewidth]{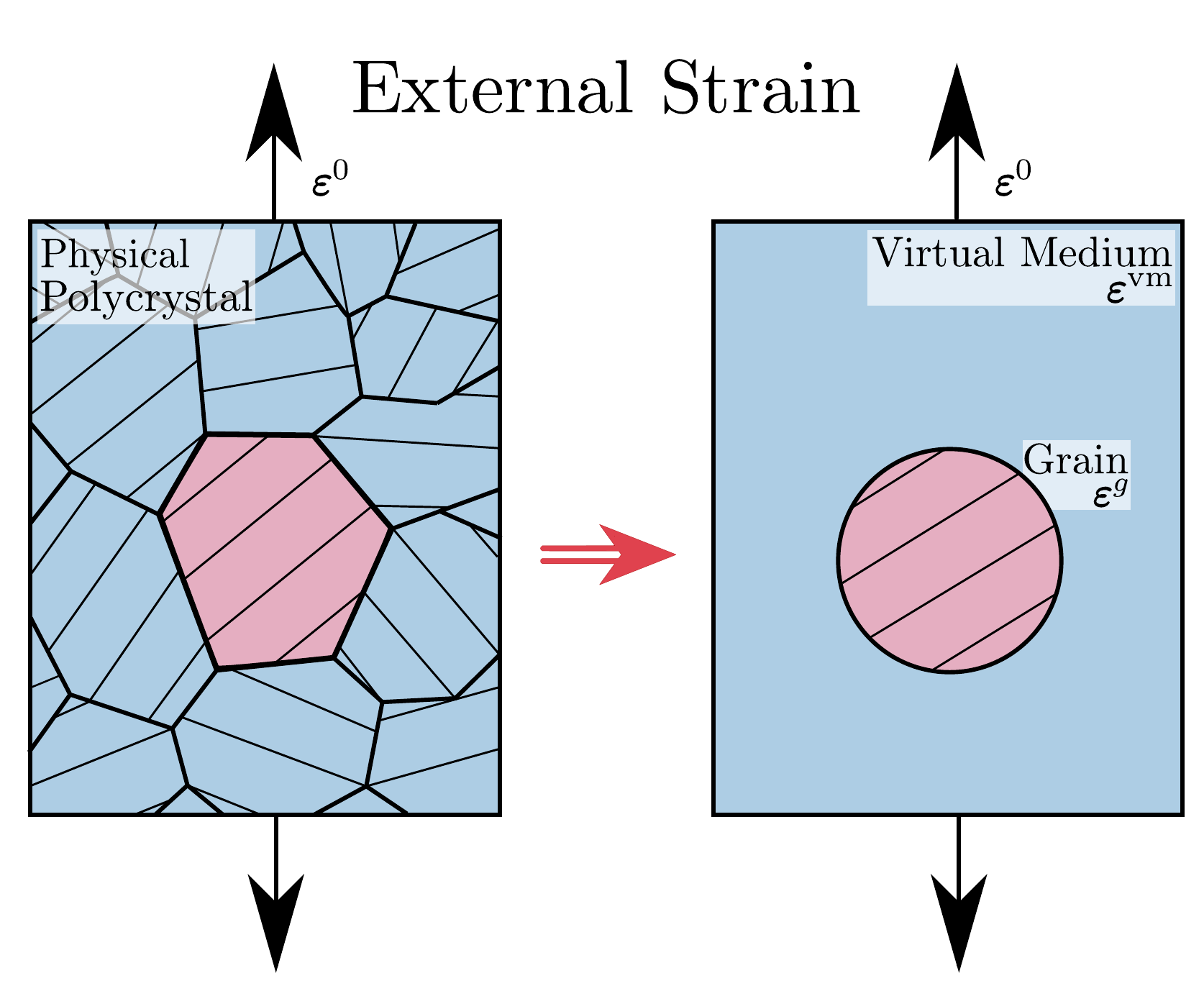}
  \caption{\label{fig:grain_inside} The theoretical model considered in Sec.~\ref{sec:InsideStress}: external strain $\bm{\varepsilon}^0$ is applied on a polycrystal, which induces strain $\bm{\varepsilon}^\text{g}$ inside a single grain. The surrounding media is assumed to be the virtual medium introduced in Sec.~\ref{sec:virtual_medium}.}
\end{figure}

In this section we calculate the stress inside a grain with arbitrary orientation, in an untextured polycrystal subjected to some external homogeneous strain $\bm \varepsilon^0$ (left panel of Fig.~\ref{fig:grain_inside}). To be able to perform an analytical calculation we assume that (i) the grain is spherical (which in average is true in the studied case) and (ii) the surrounding grains can be in average replaced by the virtual medium introduced in the previous section (right panel of Fig.~\ref{fig:grain_inside}). The latter assumption is motivated by the fact that in the untextured case the number of surrounding grains is relatively large (10-14 on average, by geometrical consideration) and is estimated by many diffraction measurements, see in Ref.~\cite{Ungar2014251, Ungar2014264, Nyilas200425, Jakobsen889, Jakobsen20073421, Levine2006}. In this section, superscript ``vm'' and ``g'' will denote the field variables in the virtual medium and in the embedded grain, respectively.

The condition for equilibrium inside the isotropic virtual medium surrounding the grain is
\begin{equation}\label{eq:eof}
\bar{\mu} \bm{\Delta} \bm{u}^\text{vm}(\bm{r})+(\bar{\lambda} + \bar{\mu})\bm{\nabla \nabla}\bm{u}^\text{vm}(\mathbf{r})=-\bm{f}(\bm{r}),
\end{equation}
where $\bar{\lambda}$ and $\bar{\mu}$ are Lam\'{e} coefficients of the virtual medium, calculated in the previous section, $\bm{u}^\text{vm}$ is the displacement vector and $\bm{f}(\bm{r})$ is the external force density. The Green's tensor of Eq.~\eqref{eq:eof} from Ref.~\cite{landau1986theory} is
\begin{equation}
G_{ij}^\text{vm}(\bm{r})=\frac{1}{8\pi \bar{\mu}}\partial_k \partial_k r \delta_{ij}- \frac{1}{8\pi \bar{\mu}} \frac{\bar{\lambda}+\bar{\mu}}{\bar{\lambda}+2\bar{\mu}} \partial_i \partial_j r.
\end{equation}
Since external forces only arise at the boundary of the sphere, for the solution of $\bm{u}^\text{vm}(\bm{r})$ one can transform the Green's spherical volume integral to a surface integral,
\begin{equation}\label{eq:u-def}
u^\text{vm}_i(\bm{r}) =\varepsilon^0_{ij}r_j + \oint df'_k G_{ij}^\text{vm}(\bm{r}-\bm{r'})P^\text{vm}_{jk}(\bm{r'}),
\end{equation}
where $\bm{\varepsilon}^0$ is the external strain applied on the polycrystal and $\bm{P}^\text{vm}$ denotes the so-called polarization tensor at the boundary of the virtual medium which depends on the local stress and displacement (for its actual form see below). The integral of Eq.~\eqref{eq:u-def} is performed on the outer surface of the grain.

To solve Eq.~\eqref{eq:u-def}, we first assume that the polarization tensor $\bm P^\text{vm}$ is constant along the surface. This assumption will be later confirmed by Eq.~\eqref{eq:P_vm}. Let us introduce the notation
\begin{equation}\label{eq:I_definition}
I_i(\bm r)=\oint df'_i \,|\bm{r}-\mathbf{r'}|=\begin{cases}
    V_a \left(\frac{a^2}{5r^2}-1 \right) \frac{r_i}{r},\text{ if $r\geq a$},\\
    V_a \left(\frac{r^3}{5a^3}-\frac{r}{a} \right) \frac{r_i}{r},\text{ if $r<a$},
  \end{cases}
\end{equation}
where $a$ is the radius of the grain, $V_a=4\pi a^3/3$ and the origin was placed in the center of the spherical grain. In the case of $r \ge a$, that is, for the virtual medium, from Eq.~\eqref{eq:u-def} one arrives at
\begin{equation}
\label{eq:u_vm}
u^\text{vm}_i(\bm r)=\varepsilon^0_{ij} r_j + \frac{P^\text{vm}_{jk}}{8\pi \bar{\mu}} \left( \partial_l \partial_l I_k \delta_{ij} - \frac{\bar{\mu}+\bar{\lambda}}{2\bar{\mu}+\bar{\lambda}} \partial_i\partial_j I_k \right).
\end{equation}
By calculating the derivatives of $I_i$, the displacement can be evaluated analytically. At the boundary ($r=a$), the displacement reads as
\begin{equation}\label{eq:insidestrain}
u^\text{vm}_i(\bm{r}) \bigg|_{r=a}=\left[\varepsilon^0_{ij} + \frac{(8\bar{\mu}+3\bar{\lambda})P^\text{vm}_{ij} - (\bar{\mu}+\bar{\lambda})\delta_{ij}P^\text{vm}_{kk}}{15\bar{\mu}(2\bar{\mu}+\bar{\lambda})} \right]r_j.
\end{equation}
From Eq.~\eqref{eq:u_vm} the strain and stress can also be calculated analytically at an arbitrary position in the virtual medium using $\varepsilon^\text{vm}_{ij}=(\partial_iu^\text{vm}_j+\partial_ju^\text{vm}_i)/2$ and
\begin{equation}\label{eq:isotrope_sigma}
\sigma^\text{vm}_{ij}=2\bar\mu\varepsilon^\text{vm}_{ij}+\bar\lambda\varepsilon^\text{vm}_{kk}\delta_{ij}.
\end{equation}
After performing the straightforward calculation one obtains at $r=a$:
\begin{equation}\label{eq:stress_vm}
\begin{split}
&\sigma^\text{vm}_{ij}(\bm{r}) \Big|_{r=a} = 2\bar\mu\varepsilon^0_{ij}+\bar\lambda\varepsilon^0_{kk}\delta_{ij} \\
&\quad +\frac{2(9\bar{\mu}+4\bar{\lambda})P^\text{vm}_{ij} + (3\bar{\lambda}-2\bar{\mu})P^\text{vm}_{kk}\delta_{ij}}{15(2\bar{\mu}+\bar{\lambda})} \\
&\quad - \frac{P^\text{vm}_{ik}r_j r_k + P^\text{vm}_{jk}r_i r_k}{a^2} \\
&\quad - \frac{2(\bar{\mu}+\bar{\lambda}) P^\text{vm}_{lk}r_i r_j r_k r_l + \bar{\lambda} a^2 P^\text{vm}_{kl} r_k r_l \delta_{ij} }{(2\bar{\mu}+\bar{\lambda}) a^4}.
\end{split}
\end{equation}

Now we continue with employing the boundary conditions at the grain surface. First, the value of the displacement $\bm u$ must be equal on the two sides of the boundary ($r=a$):
\begin{equation}
\begin{split}
&u^\text{g}_i(\bm{r})\Big|_{r=a} = u^\text{vm}_i(\bm{r})\Big|_{r=a} \\
&\quad =\left[\varepsilon^0_{ij} + \frac{(8\bar{\mu}+3\bar{\lambda})P^\text{vm}_{ij} - (\bar{\mu}+\bar{\lambda})\delta_{ij}P^\text{vm}_{ll}}{15\bar{\mu}(2\bar{\mu}+\bar{\lambda})} \right]r_j.
\end{split}
\label{eq:u_boundary}
\end{equation}
Secondly, the normal component of the stress tensor must also be equal on the two sides, which can be formulated as
\begin{equation}\label{eq:boundary_sigma}
\sigma^\text{g}_{ij} r_j \Big|_{r=a} = \sigma^\text{vm}_{ij} r_j \Big|_{r=a}
\end{equation}
since $\bm r$ is always perpendicular to the spherical surface. After inserting the analytical expression for $\bm \sigma^\text{vm}$ [Eq.~\eqref{eq:stress_vm}] into Eq.~\eqref{eq:boundary_sigma} one obtains
\begin{equation}
\begin{split}
&\sigma^\text{g}_{ij} r_j \Big|_{r=a} = \Bigg[ 2\bar{\mu}\varepsilon^0_{ij}+ \bar{\lambda}\varepsilon^0_{kk}\delta_{ij} \\ 
 & \quad +\frac{(3\bar{\lambda}-2\bar{\mu})P^\text{vm}_{kk}\delta_{ij}-(14\bar{\mu}+9\bar{\lambda})P^\text{vm}_{ij}}{15(2\bar{\mu}+\bar{\lambda})} \Bigg] r_j.
\end{split}
\label{eq:sigma_r}
\end{equation}

In the studied case the inclusion is spherical and the external applied strain is homogeneous, therefore, based on Eshelby's theorem \cite{Eshelby376,sneddon1961progress} one concludes that both strain and stress ($\bm{\varepsilon}^\text{g}$ and $\bm{\sigma}^\text{g}$, respectively) are homogeneous inside the grain. The condition on strain means that the displacement inside the grain is of the form $u_i^\text{g} = \varepsilon_{ij}^\text{g} r_j$. Comparing this with Eq.~\eqref{eq:u_boundary} it follows that
\begin{equation}\label{eq:eps}
\varepsilon^\text{g}_{ij}= \varepsilon^0_{ij} + \frac{(8\bar{\mu}+3\bar{\lambda})P^\text{vm}_{ij} - (\bar{\mu}+\bar{\lambda})\delta_{ij}P^\text{vm}_{kk}}{15\bar{\mu}(2\bar{\mu}+\bar{\lambda})}.
\end{equation}
According to the condition on stress, $\sigma_{ij}^\text{g}$ is independent of $\bm r$ in Eq.~\eqref{eq:sigma_r}. Since the expression in the brackets on the right-hand-side is also constant throughout the boundary it follows that
\begin{equation}\label{eq:sigma}
\begin{split}
\sigma^\text{g}_{ij} &= 2\bar{\mu}\varepsilon^0_{ij}+ \bar{\lambda}\varepsilon^0_{kk}\delta_{ij} \\ 
 & \quad +\frac{(3\bar{\lambda}-2\bar{\mu})P^\text{vm}_{kk}\delta_{ij}-(14\bar{\mu}+9\bar{\lambda})P^\text{vm}_{ij}}{15(2\bar{\mu}+\bar{\lambda})}.
 \end{split}
\end{equation}

Our aim is to express the internal strain $\bm \varepsilon^\text{g}$ in terms of the applied strain $\bm \varepsilon^0$. To this end, we first eliminate the polarization tensor $\bm P^\text{vm}$ from the equations by first taking the trace of Eq.~\eqref{eq:eps} leading to
\begin{equation}
P^\text{vm}_{ii}=3(2\bar{\mu}+\bar{\lambda})\Delta\varepsilon_{ii},
\end{equation}
where the notation $\Delta\varepsilon_{ij}=\varepsilon^\text{g}_{ij}-\varepsilon^0_{ij}$ is introduced. From Eq.~\eqref{eq:eps} $\bm P^\text{vm}$ can be expressed with $\Delta\bm\varepsilon$ as
\begin{equation}
\label{eq:P_vm}
P^\text{vm}_{ij}=\frac{3(2\bar{\mu}+\bar\lambda)}{8\bar\mu+3\bar\lambda}\left[5\bar\mu\Delta\varepsilon_{ij}+(\bar\mu+\bar\lambda)\Delta\varepsilon_{kk}\delta_{ij}\right].
\end{equation}
Inserting this expression into Eq.~\eqref{eq:sigma} and using the relation
\begin{equation}\label{eq:grain_stress_strain}
\sigma^\text{g}_{ij}= C_{ijkl}\varepsilon^\text{g}_{kl},
\end{equation}
($C_{ijkl}$ is the stiffness tensor of the anisotropic grain) one obtains the final result
\begin{equation}\label{eq:solution}
 C_{ijkl}\varepsilon^\text{g}_{kl}+A_1\varepsilon^\text{g}_{ij}+A_2\varepsilon^\text{g}_{kk}\delta_{ij}= (2\bar{\mu}+A_1)\varepsilon^0_{ij}+(\bar{\lambda}+A_2)\varepsilon^0_{kk}\delta_{ij},
\end{equation}
where $A_1=\bar{\mu}(14\bar{\mu}+9\bar{\lambda})/(8\bar{\mu}+3\bar{\lambda})$ and $A_2=\bar{\mu}(6\bar{\mu}+\bar{\lambda})/(8\bar{\mu}+3\bar{\lambda})$. Equation~\eqref{eq:solution} is the desired formula that provides the strain inside the anisotropic grain as a function of the applied strain and the elastic coefficients of the material.

The method for obtaining the strain $\bm \varepsilon^\text{g}$ is, thus, as follows: With the help of anisotropic $C_{ijkl}$ coefficients of the material, the virtual $\bar{\mu}$ and $\bar{\lambda}$ Lam\'{e} coefficients can be calculated from Eq.~\eqref{eq:geomean}. Then, at a given external strain $\bm \varepsilon^0$ the internal strain can be obtained from Eq.~\eqref{eq:solution}. The internal stress field follows from Eq.~\eqref{eq:grain_stress_strain}.

\begin{figure}[h]
\centering
  \includegraphics[width=\linewidth]{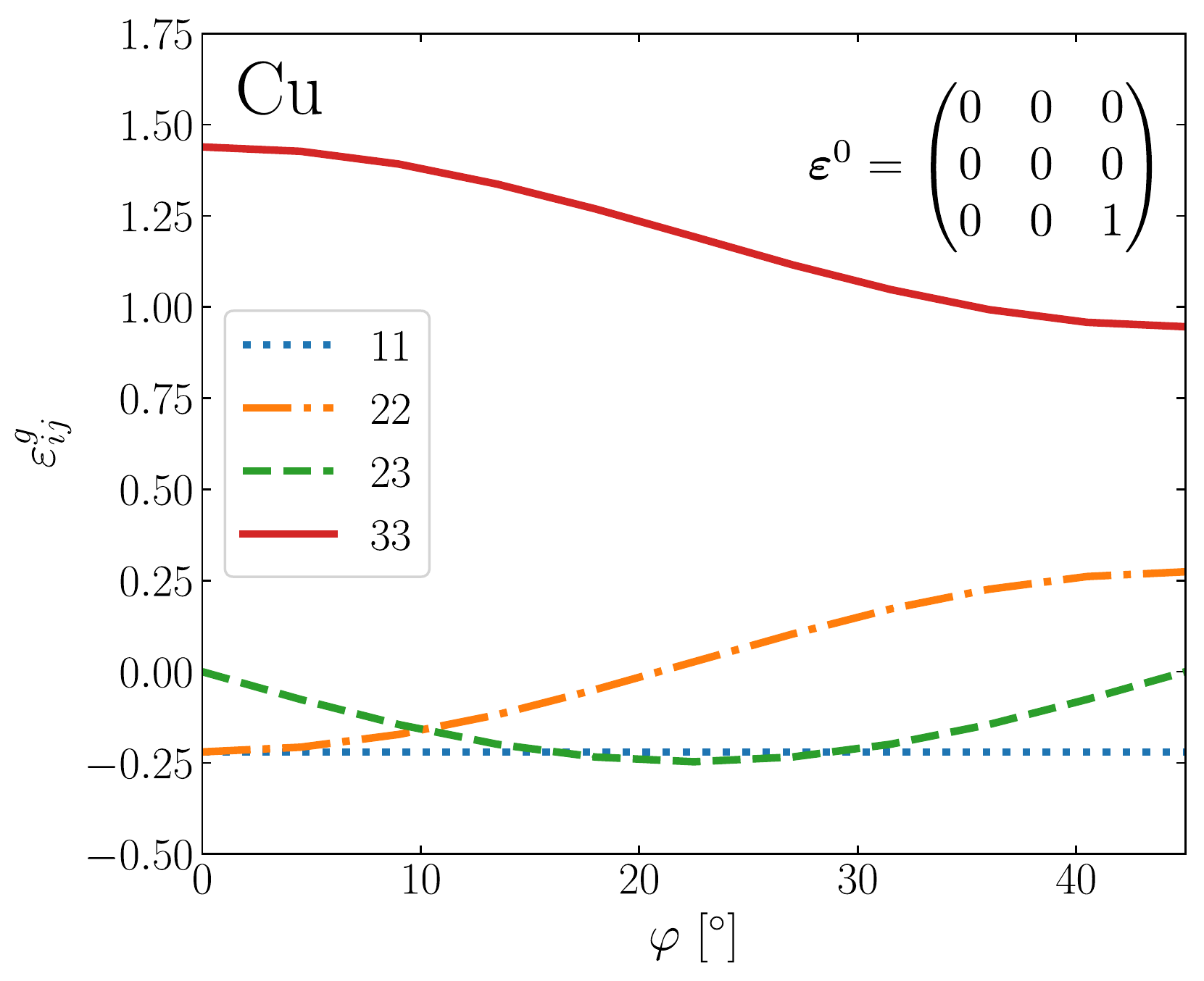}
  \includegraphics[width=\linewidth]{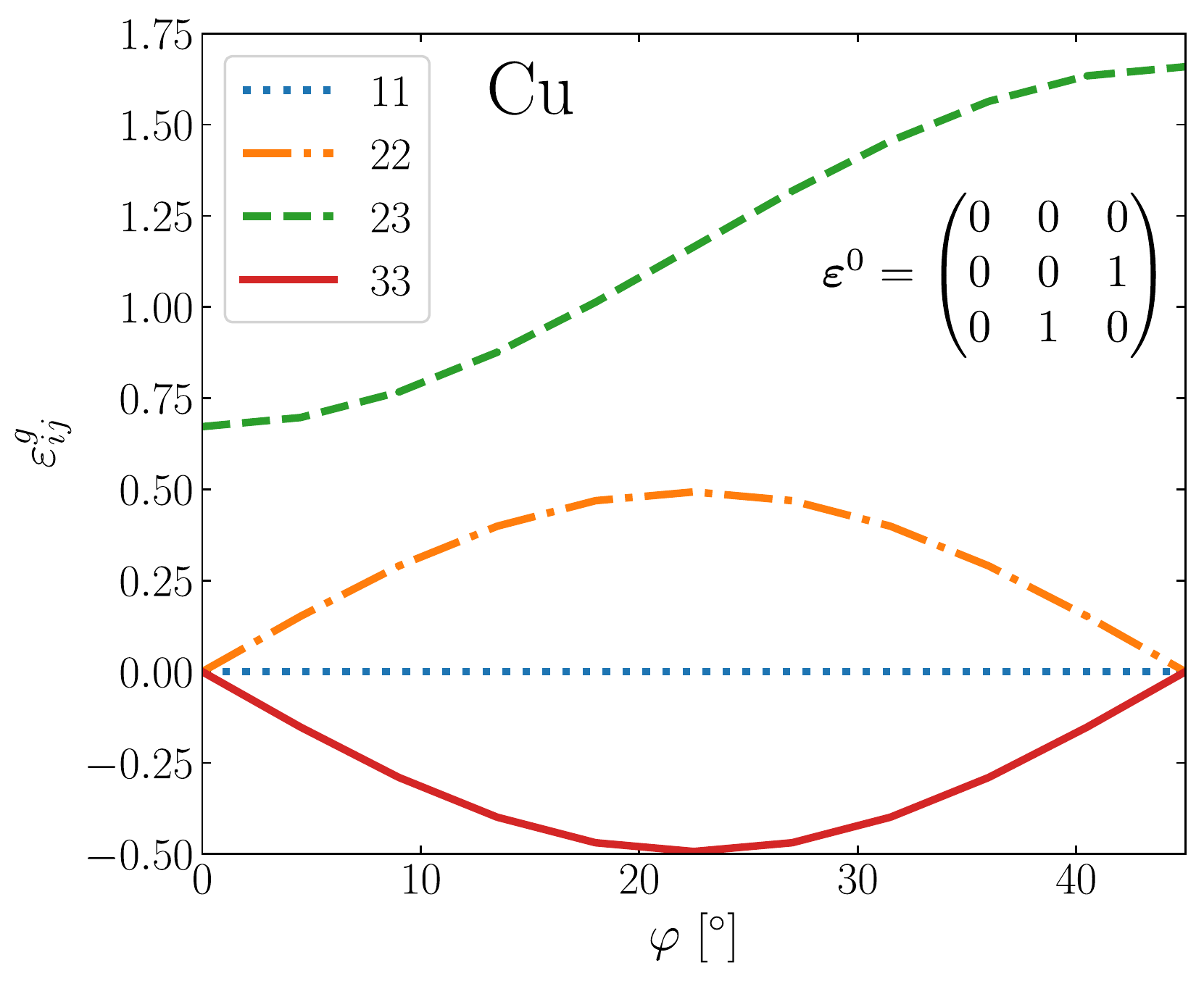}
  \caption{\label{fig:cu}The strain $\bm \varepsilon^\text{g}$ inside a Cu grain, as the function of relative angle $\varphi$ between the orientation of the grain and the loading direction (see details in the text). In the top panel pure tension is considered while bottom panel depicts pure external shear.}
\end{figure}

To illustrate the result, we used Eq.~\eqref{eq:solution} to calculate the orientation dependence of the internal strain $\bm\varepsilon^\text{g}$ at a given homogeneous external strain $\bm\varepsilon^0$. The coordinate system is attached to the virtual medium and the grain of cubic symmetry is rotated around the $x$ axis with different angles $\varphi$. In this example we used the elastic coefficients of copper from Tab.~\ref{tab:coeff1}. First, we considered tension along the $z$ axis ($\varepsilon^0_{33}=1)$, see the upper panel of Fig.~\ref{fig:cu}. In accordance with one's expectations strain component parallel to the tensile axis $\varepsilon_{33}$ decreases with increasing angle $\varphi$ whereas the lateral deformation component $\varepsilon_{22}$ increases. It is also seen that $\varepsilon_{11}$ is constant because the rotation takes place around the $x$-angle and that rotation induces shear strain in the grain. Secondly, we consider the case of pure shear ($\varepsilon^0_{23}=\varepsilon^0_{32}=1$), depicted in the lower panel of Fig.~\ref{fig:cu}. Again, the observed behavior of strain values is in line with physical expectations.

In the next section, in order to verify this model we continue with comparing its predictions with experimental results.

\section{Application for Diffraction Measurements}\label{sec:application}

\begin{figure}[h]
\centering
    \includegraphics[width=\linewidth]{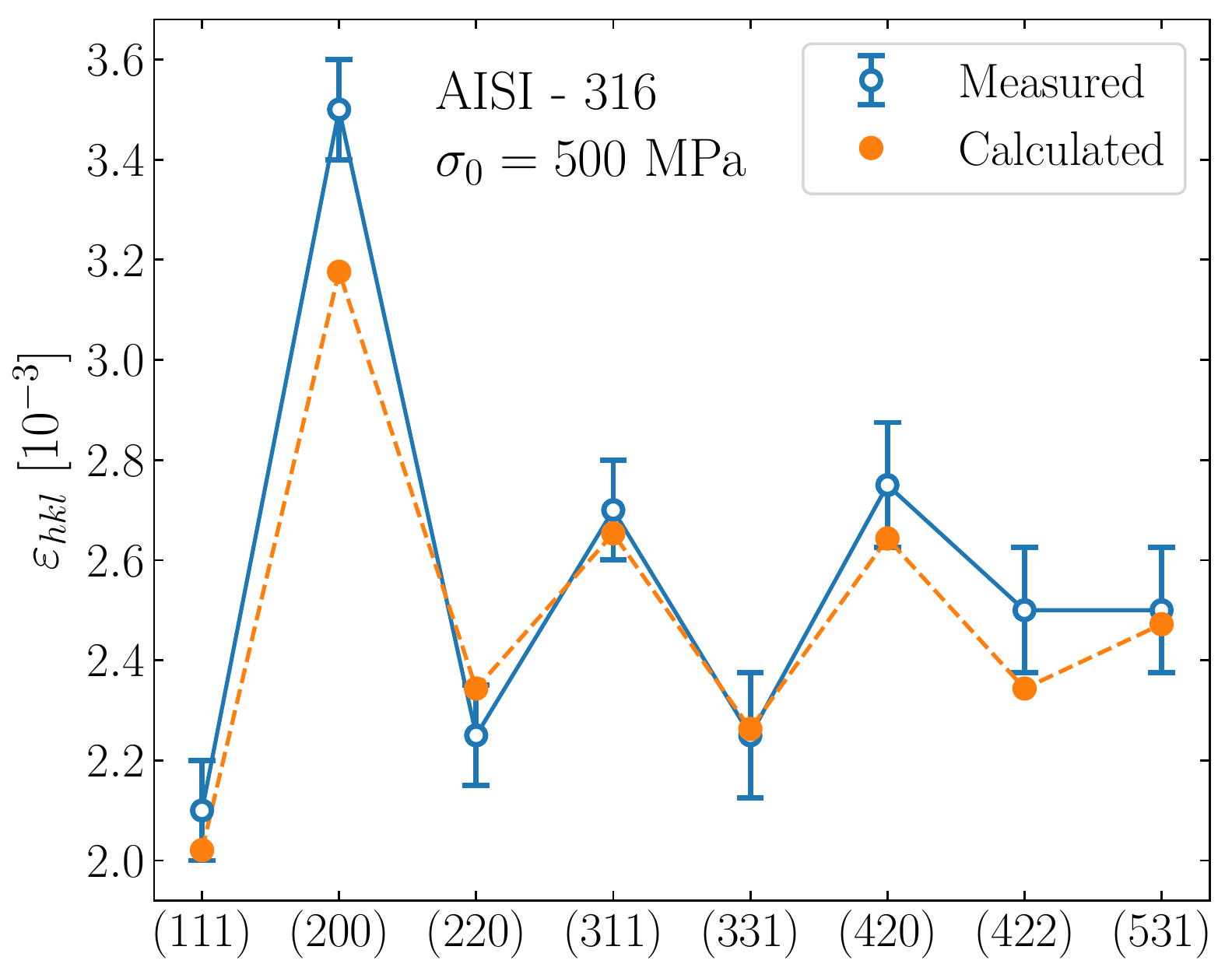}
    \caption{\label{fig:aisi-316} The measured and calculated relative elongation of the reflection planes with different $(hkl)$ indices. The measurement was performed on AISI-316 stainless polycrystal steel, see Ref.~\cite{Ungar2014264}.}
\end{figure}

In Ref.~\cite{Ungar2014264} the authors performed neutron diffraction measurements on an untextured 316 steel polycrystal. First, they measured diffraction patterns without applied load to determine the diffraction planes and the initial value of the lattice constants. Then the measurement was repeated with an externally applied homogeneous tensile load below the yielding threshold. From the shifts of the Bragg reflections, the axial lattice strain $\varepsilon_{hkl}$ was determined at several Miller indices. Since for a given $hkl$ index only grains with orientation satisfying the Bragg condition contribute to the diffraction pattern, the measured strain corresponds to that orientation. The results are shown in Fig.~\ref{fig:aisi-316} in blue. The applied stress was 500 MPa and the error bars represent the uncertainties of the measurement.
Now we make an estimation on the same strain values using the theory introduced previously. The elastic coefficients of the steel from Ref.~\cite{BAUDOUIN201393} are
\begin{equation}
C_{11}=210 \text{ GPa}, \, C_{12}=130 \text{ GPa},\, C_{44}=120 \text{ GPa}.
\end{equation}
The calculated isotropic virtual coefficients are $\bar{\mu}=76.6$ GPa and $\bar{\lambda}=106$ GPa. The external strain is determined using 
\begin{equation}\label{eq:appl_external_strain}
\varepsilon^0_{ij}=(\overline{\bm{S}}_{av})_{ijkl}\sigma^0_{kl} ,
\end{equation}
where $\overline{\bm{S}}_{av}$ is determined by the virtual medium from Eq.~\eqref{eq:geomean}. The applied true stress during the measurement was homogeneous $\sigma^0=500$ MPa and we choose the $z$ axis to be parallel with the loading axis. The reciprocal lattice vector is $\bm{g}=(h,k,l)/a$, where $a$ is the lattice constant. Inserting Eq.~\eqref{eq:appl_external_strain} in the implicit Eq.~\eqref{eq:solution}, the measured relative displacement can be calculated analytically from
\begin{equation}\label{eq:lattice_strain}
\varepsilon_{hkl}=\left. \frac{g_i \varepsilon^\text{g}_{ij}g_j}{g^2} \right|_{hkl},
\end{equation}
where, $\varepsilon^{\text{g}}_{ij}$ depends on $\sigma^0,C_{ijkl},\bar{\lambda}$ and $\bar{\mu}$. The calculated values of Eq.~\eqref{eq:lattice_strain} are shown in Fig.~\ref{fig:aisi-316} for reflections with different Miller indices. According to the comparison with the measured values the deviation is only around a few percent, supporting the assumptions of our model. The small deviations observed are probably due to the material not being perfectly untextured.

\section{Summary}

In this paper a new method for calculating elastic stresses and strains in untextured polycrystalline materials was presented. First, a technique for directional averaging on tensors of any rank was reviewed followed by the determination of elastic moduli of a bulk untextured polycrystal. The theory was applied for crystals of cubic symmetry and the predicted elastic constants showed good agreement with measurements. Secondly, an analytical formula was derived that describes the average internal stress and strain within individual grains of a polycrystal of different crystallographic orientation. The derivation was based on the following assumptions:
\begin{itemize}
\item[(i)] The grains are considered spherical;
\item[(ii)] The material surrounding the grain with specific orientation was assumed to be isotropic with elastic constants calculated in the first part of the paper; 
\item[(iii)] The displacement $\bm u$ is considered continuous at the grain boundary. This assumes pure elasticity, that is, a dislocation free boundary. Upon significant plastic deformation this condition may be violated~\citep{2002JMPSo..50.2597K,KNEZEVIC20132034,Lebensohn2011}.
\end{itemize}
The calculations yielded Eq.~\eqref{eq:solution}, which is an analytical formula that provides the strain inside the considered grain as a function of the applied load.

The assumptions listed above may seem strict but, as it was shown in Sec.~\ref{sec:application}, the theory can be directly applied for untextured polycrystalline materials. Neutron diffraction experiments were analyzed and we calculated the strain in grains with identical orientation, from the shift of the corresponding Bragg peaks. The comparison of these data with ones predicted by the theory showed excellent agreement validating the model assumptions.

The advantage of the model presented mostly lies in its simplicity, because it is capable of giving accurate predictions about internal elastic stresses and strains without the detailed knowledge of the microstructure. In addition, solution of the derived equation is orders of magnitude faster than that of a sophisticated simulation tool (e.g., CPFE).

The main area of application for the presented theory is expected to be the evaluation of diffraction experiments, as demonstrated in Sec.~\ref{sec:application}. In particular, during plastic deformation average stress and strain values within individual grains can be estimated. This theory is also expected to facilitate developing plasticity theories of polycrystalline materials.

The presented calculations are performed for untextured materials, however, they can be generalized in a straightforward manner for a textured microstructure as it is detailed below. When calculating the elastic constants of the virtual medium the texture could be accounted for in terms of using specific weight-factors in Eq.~\eqref{eq:rotational_average}. The virtual medium would then naturally lose its isotropic character. In the elastic calculation of Sec.~\ref{sec:InsideStress} shape of grains could be taken into account by considering the grain to be ellipsoidal. This would still permit the applicability of Eshelby's theorem and would give a good approximation for the average grain shape for most cases. So, the analytical derivation could be easily generalized allowing a wider applicability of the present results. This calculation is relegated for future work.

\begin{acknowledgments}
We would like to express our appreciation to Prof. Jen\H{o} Gubicza for his advices. \'{A}T is supported by the \'UNKP-18-3 New National Excellence Program of the Hungarian Ministry of Human Capacities. PDI is supported by the \'UNKP-18-4 New National Excellence Program of the Hungarian Ministry of Human Capacities and by the J\'anos Bolyai Scholarship of the Hungarian Academy of Sciences. This work was completed in the ELTE Institutional Excellence Program (1783-3/2018/FEKUTSRAT) supported by the Hungarian Ministry of Human Capacities. Financial support of the National Research, Development and Innovation Found of Hungary (project Nos. NKFIH-K-119561 and NKFIH-KH-125380) is gratefully acknowledged.
\end{acknowledgments}

\bibliography{references}

\end{document}